\documentclass[aps,prl,twocolumn,showpacs,superscriptaddress,floatfix]{revtex4-1} 
\usepackage{amsmath,amssymb}
\usepackage{graphicx}
\usepackage[pdftex,colorlinks,citecolor=blue,linkcolor=blue,bookmarks=false,urlcolor=blue]{hyperref} %
\usepackage{amsfonts,bm}

\newcommand{\ot}{\otimes}
\newcommand{\bea}{\begin{eqnarray}}
\newcommand{\eea}{\end{eqnarray}}
\newcommand{\se}[1]{\noindent{\bf #1 }}

\newcommand{\cl}[1]{{q_{#1}}}
\newcommand{\opl}{+}
\newcommand{\ii}{\mathrm{i}}
\newcommand{\re}{\mathrm{e}}
\newcommand{\bbn}{\mathbb{N}}
\newcommand{\ov}[1]{\overline{#1}}
\newcommand{\GSD}{\text{GSD}}
\newcommand{\Tr}{\mathop{\mathrm{Tr}}}
\newcommand{\Diag}{\mathop{\mathrm{Diag}}}
\newcommand{\one}{\bm 1}

\newcommand{\cS}{\mathcal{S}}
\newcommand{\tS}{\tilde{\cS}}
\newcommand{\cT}{\mathcal{T}}
\newcommand{\cK}{\mathcal{K}}
\newcommand{\cL}{\mathcal{L}}
\newcommand{\cM}{\mathcal{M}}
\newcommand{\cN}{\mathcal{N}}
\newcommand{\cW}{\mathcal{W}}
\newcommand{\iden}{I}

\newcommand{\del}[1]{{\iffalse #1 \fi}}
\newcommand{\add}[1]{#1}

\begin{document}

\title{Gapped Domain Walls, Gapped Boundaries and Topological Degeneracy}
\author{Tian Lan} \email{tlan@perimeterinstitute.ca}
\affiliation{Perimeter Institute for Theoretical Physics, Waterloo, ON, Canada N2L 2Y5}
\affiliation{Department of Physics and Astronomy,
University of Waterloo, Waterloo, ON, Canada N2L 3G1}
\author{Juven C. Wang} \email{juven@mit.edu}
\affiliation{Department of Physics, Massachusetts Institute of Technology, Cambridge, MA 02139, USA}
\affiliation{Perimeter Institute for Theoretical Physics, Waterloo, ON, Canada N2L 2Y5}
\author{Xiao-Gang Wen} \email{wen@dao.mit.edu}
\affiliation{Perimeter Institute for Theoretical Physics, Waterloo, ON, Canada N2L 2Y5}
\affiliation{Department of Physics, Massachusetts Institute of Technology, Cambridge, MA 02139, USA}
\affiliation{Institute for Advanced Study, Tsinghua University, Beijing, 100084, P. R. China}

\begin{abstract}

Gapped domain walls, as topological line defects between 2+1D topologically
ordered states, are examined. 
We provide simple criteria to determine the existence of gapped domain walls,
which apply to both Abelian and non-Abelian topological orders.
Our criteria also determine which 2+1D topological orders must have gapless
edge modes, namely which 1+1D \emph{global gravitational anomalies} ensure gaplessness.
Furthermore, we introduce a new mathematical object, the \emph{tunneling matrix}
$\cW$, whose entries are the \emph{fusion-space dimensions} $\cW_{ia}$, to
label different types of gapped domain walls. 
By studying many examples,
we find evidence that the tunneling matrices are powerful quantities to
classify different types of gapped domain walls.  
Since a gapped boundary is a gapped domain wall between a bulk topological
order and the vacuum, regarded as the trivial topological order, our theory of
gapped domain walls inclusively contains
the theory of gapped boundaries.
In addition, we derive a topological ground state degeneracy formula, applied to
arbitrary orientable spatial 2-manifolds with gapped domain walls, including
closed 2-manifolds and open 2-manifolds with gapped boundaries.

\end{abstract}
\maketitle
\se{Introduction}--
Insulator has a finite energy gap, which is rather trivial at low energy.
Nonetheless, domain walls, separating different symmetry-breaking insulating regions, can enrich the physics of a trivial insulator, such as some paramagnet~\cite{ashcroft1976solid}.
\emph{Topological order}~\cite{Wen89,WN90,Wen90}, on the other hand, as a new kind of many-body quantum ordering,  
has a gapped bulk with exotic properties:  
some have (i) gapless edge modes, (ii) 
anyonic excitations with fractional or non-Abelian statistics~\cite{2008RMPNab}, such as fractional quantum Hall states, and (iii)
long-range entanglement~\cite{KP0604,LevinWen2006,ChenGuWen2010}.
In this Letter, we would like to investigate the gapped domain walls of topological orders, and how gapped domain walls further enrich their physics. 

It was conjectured that the 2+1D topological orders
are completely classified by the 
gauge connection on the moduli space of the degenerate
ground states~\cite{Wen90,KW9327}.
The non-Abelian part of the  gauge connection is the
non-Abelian geometric phase~\cite{Wilczek1984dh} characterized by the $\cS,\cT$ matrices, which also encode the anyon statistics.
The Abelian part is related to the
gravitational Chern-Simons term in the effective theory
and is described by the chiral central charge $c_-$ of the edge state.
Non-zero $c_-$ implies robust gapless edge modes.

By now we understand how to label a 2D
topological order
by a set of ``\emph{topological order parameters}'' ($\cS,\cT,c_-$),
analogous to ``symmetry-breaking {order parameters}'' for spontaneous symmetry breaking systems~\cite{GL5064,LanL58}.  
However, it is less known how different topological orders are related.
To this end, it is important to investigate the following circumstance: there are several
domains in the system and each domain contains a topological order, while
the whole system is gapped. In this case, different topological orders are
connected by \emph{gapped domain walls}. Our work addresses two primary questions: \\
\noindent
{\bf (Q1)} ``\emph{Under what criteria can two topological orders be
connected by a gapped domain wall, and how many different types of gapped domain walls are there}?''
Since a gapped boundary is a
gapped domain wall between a nontrivial topological order and the vacuum,
we also address that
``\emph{under what criteria can topological orders allow gapped boundaries}?'' 

\noindent 
{\bf (Q2)} ``\emph{When a topologically ordered system has a gapped bulk, gapped domain walls and gapped boundaries,
how to calculate its ground state degeneracy} 
(GSD)~\cite{Wen89,WN90,WW12,Kap14,CWangLevin1311}, \emph{on any orientable manifold?}''

\se{Main result}--
Consider two topological orders, Phases $A$ and $B$, described by $(\cS^A,
\cT^A,c^A_-)$ and $(\cS^B,\cT^B,c^B_-)$.
Suppose there are $N$ and $M$ types of anyons in Phase $A$ and Phase $B$,
then the ranks of their modular matrices are $N$ and $M$ respectively.
If $A$ and $B$ are connected by a gapped domain wall, firstly their central
charges must be the same $c^A_-=c^B_-$.
Next, we find that the domain wall can be labeled by
a $M\times N$ \emph{tunneling matrix} $\cW$ whose entries are \emph{fusion-space dimensions} $\cW_{ia}$
satisfying the \emph{commuting condition} \eqref{commute},
and the \emph{stable condition}~\eqref{stable}:
\begin{gather}
\cW_{ia}\in\bbn,
 \label{Winteger}\\
  \cS^B \cW=\cW \cS^A,\quad \cT^B \cW= \cW \cT^A,
  \label{commute}\\
  \cW_{ia}\cW_{jb}\leq\sum_{kc} (\cN^B)_{ij}^k \cW_{kc} (\cN^A)_{ab}^c\,.
  \label{stable}
\end{gather}
$\bbn$ denotes the set of non-negative integers. 
$a,b,c,\dots$ and $i,j,k,\dots$ are anyon indices for Phases $A,B$.
$(\cN^A)_{ab}^c$ and $(\cN^B)_{ij}^k$ are fusion tensors~\cite{Verlinde1988sn,2008RMPNab} of Phases $A,B$.

\eqref{Winteger}\eqref{commute}\eqref{stable} is a set of necessary conditions a gapped domain wall must satisfy,
i.e., \emph{if there is no non-zero solution of $\cW$, the domain wall must be gapless.}
We conjecture that they are also sufficient \add{for a gapped domain wall to exist.}
In the examples studied in Supplemental Material, $\cW$ are in one-to-one correspondence with gapped domain walls.
However, for some complicated examples~\cite{Dav13}, a $\cW$ matrix may correspond to more than one type of gapped domain wall.
This indicates that some additional data are needed to completely classify gapped domain walls.

\begin{figure}
  \includegraphics[scale=0.75]{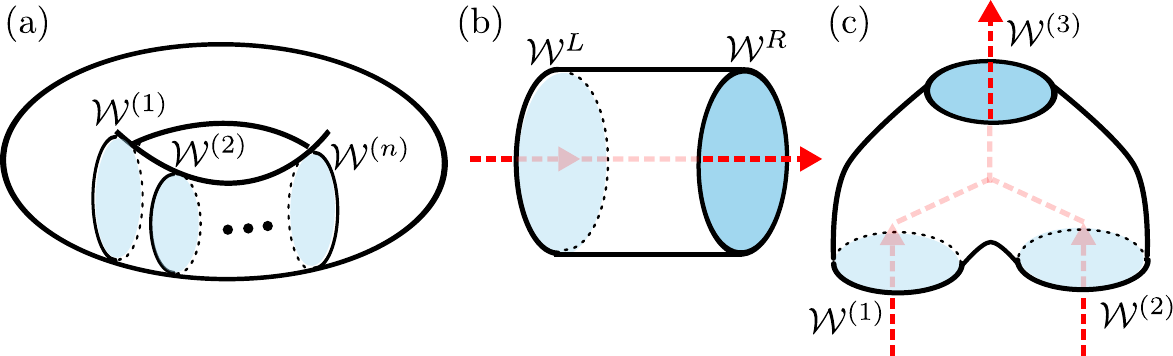}
  \caption{Some 2-manifolds with gapped domain walls. 
  }
  \label{fig:gdwfigure}
\end{figure}

As a first application of our result, we give a general method to compute the
GSD in the presence of gapped domain walls on any orientable 2D surface.  A
simple case is the GSD on a disk drilled with two holes (equivalently a sphere
with 3 circular boundaries, see Fig.~\ref{fig:gdwfigure}(c)). The gapped
boundaries are labeled by three \emph{vectors}
(one-row or one-column matrices) $\cW^{(1)},\cW^{(2)},\cW^{(3)}$. The GSD is
${\sum}_{ijk} \cW^{(1)}_{i1} \cW^{(2)}_{j1} \cN^k_{ij} \cW^{(3)}_{1k}$.

For gapped boundaries,
our criteria can be understood via \emph{dimension reduction},
i.e., shrinking a 1D gapped boundary $\cW$ to a (composite \footnote{The concepts of trapping anyons,
composite anyon types and fusion spaces are discussed in~\cite{LW14}}) anyon $\bm
q_\cW = \oplus_a\cW_{1a}{a}$.
If the system is on a 2D surface $M^2$ drilled with $n$ gapped boundaries ${\cW^{(1)}}, \dots, {\cW^{(n)}}$, then the GSD is the dimension of the fusion space~\cite{Note1} 
with anyons $\bm{q}_{\cW^{(1)}},  \dots, \bm{q}_{\cW^{(n)}}$,
$\GSD=\dim[\mathcal V(M^2,\bm{q}_{\cW^{(1)}},\dots, \bm{q}_{\cW^{(n)}}  ) ].  $

Since gapped domain walls \emph{talk to each other} through long-range
entanglement, the GSD with domain walls reveals more physics than that without
domain walls.  We foresee its practicality in experiments, since we can read
even more physics by putting the system on open surfaces with gapped domain
walls.
Below we shall properly introduce $\cS,\cT$ and $\cW$ matrices.

\se{Modular $\cS,\cT$ matrices}--
$\cS$ and $\cT$ are unitary matrices indexed by anyon types $\{1,a,b,c,\dots\}$. 1 labels the trivial anyon type.
The anti-quasiparticle of $a$ is denoted by $a^*$.

$\cT$ describes the self statistics.
It is diagonal $\cT_{ab}=\re^{\ii\theta_a}\delta_{ab}$, where $\re^{\ii\theta_a}$ is the phase factor when
exchanging two anyons $a$.
For the trivial type, 
$\cT_{11}=\re^{\ii\theta_1}=1$.
$\cS$ describes the mutual statistics.
$\cS_{ab}$ is the amplitude of the following process with proper normalization factors:
first create a pair of $aa^*$ and a pair of $bb^*$, then braid $a$ around $b$,
and finally annihilate the two pairs.
$\cS$ is symmetric, $\cS_{ab}=\cS_{ba}$.
If $b=1$, the process is just creation and annihilation, and $\cS_{a1}>0$.
$\cS$ and $\cT$ form a projective representation of
the modular group:
  $\cS^4=\iden,
  (\cS \cT)^3=\re^{2\pi\ii c_-/8} \cS^2$, where $\iden$ denotes the identity matrix.

The anti-quasiparticle can be read from $\cS^2$, ${(\cS^2)_{ab}=\delta_{a^*b}}$.
The fusion tensor $\cN_{ab}^c$ can be calculated via the \emph{Verlinde formula}~\cite{Verlinde1988sn}:
    \begin{align}
      \cN_{ab}^c=\sum_{m} \frac{\cS_{am} \cS_{bm}\ov{\cS_{cm}}}{\cS_{1m}}\in
      \bbn.
      \label{verlinde}
    \end{align}

\se{Gapped domain walls}--
Below we demonstrate the physical meanings of the gapped domain wall conditions \eqref{Winteger}\eqref{commute}\eqref{stable}.
First we put Phase $A$ and Phase $B$  on a sphere $S^2$, separated by a gapped
domain wall. Note that there can be many types of domain walls
separating the same pair of phases $A$ and $B$.  What data characterize those different types of domain walls?
We fix the domain wall type, labeled by $W$, and trap~\cite{Note1} an anyon $a^*$ in
Phase $A$, an anyon $i$ in Phase $B$ and. This configuration is denoted by
$(S^2,i,W,a^*)$.  The states with such a configuration may be degenerate
and the degenerate subspace is the fusion space $\mathcal V(S^2,i,W,a^*)$.
Here we propose using the
\emph{fusion-space dimensions
  $\cW_{ia} \equiv {\dim}[\mathcal V(S^2,i,W,a^*)]\in\bbn$} to characterize the
gapped domain wall $W$.
\del{Below we will replace the abstract label $W$ by the concrete data $\cW$.}

\add{There are non-local operators $O_{W,ia^*}$ that create a
pair $aa^*$ in Phase $A$, and then tunnel $a$ through the domain wall to an anyon $i$ in Phase $B$,
$O_{W,ia^*}|\psi_{S^2,W}\rangle\in\mathcal V(S^2,i,W,a^*)$,
where $|\psi_{S^2,W}\rangle$ is the ground state.
Since we care about the fusion states rather than the operators themselves,
we would take the equivalent class $[O_{W,ia^*}]=\left\{U_{W,ia^*}\middle|(O_{W,ia^*}-U_{W,ia^*})|\psi_{S^2,W}\rangle=0\right\}$.
We call $[O_{W,ia^*}]$ as \emph{tunneling channels},
which correspond to fusion states in $\mathcal V(S^2,i,W,a^*)$.
Therefore, the fusion space dimension $\cW_{ia}$ is the number of linearly independent tunneling channels.
So, we also refer to $\cW$ as the ``tunneling matrix.''
}

\del{We can compute the dimension of the  fusion space $\mathcal V(S^2,i,W,a^*)$ by first creating a
pair $aa^*$ in Phase $A$, then tunneling $a$ through the domain wall.
In the channel where the tunneling does not leave any topological quasiparticle
on the domain wall, $a$ in Phase $A$ will become
a composite anyon $\bm q_{\cW,a}=\oplus_i \cW_{ia} i$ in Phase $B$.
Thus, the fusion-space dimension $\cW_{ia}$ is also the \emph{number} of tunneling channels from,
$a$ of Phase A, to, $i$ of Phase $B$. So, we also refer to $\cW$ as
the ``tunneling matrix.''}

The \emph{commuting condition} \eqref{commute} dictates the consistency of anyon statistics in presence of gapped domain walls.
Since modular $\cS,\cT$ matrices encode the anyon statistics,
we require that $\cW$ should commute with them as \eqref{commute}:
$\cS^B \cW=\cW \cS^A$, $\cT^B \cW= \cW \cT^A$.

We may as well create a pair $ii^*$ in Phase $B$ and tunnel $i^*$ to $a^*$.
$\cW^\dag$ describes such tunneling in the opposite direction (i.e.,
$\cW:A\to B,~ \cW^\dag: B\to A$). $\cW^\dag$ and $\cW$ contains the same physical
data. To be consistent, tunneling $i^*$ to $a^*$ should give the same
fusion-space dimension, $(\cW^\dag)_{a^*i^*}=\cW_{i^*a^*}=\cW_{ia}$.
\add{This is guaranteed by $\cW (\cS^A)^2=(\cS^B)^2\cW$ and $(\cS^2)_{ab}=\delta_{a^*b}$.}

\begin{figure}
  \centering
    \includegraphics{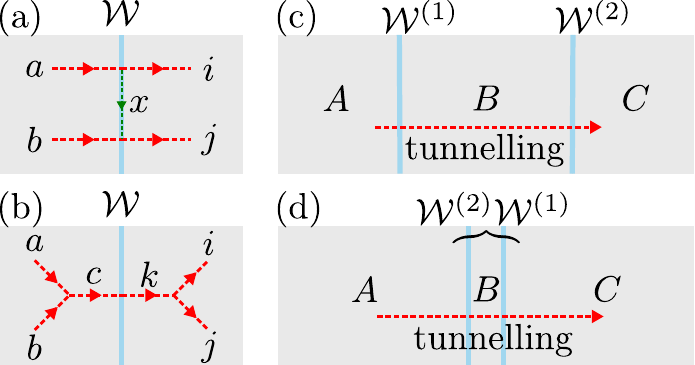}
  \caption{(a)(b) Tunneling channels. 
 (c) Separated domain walls
  $\cW^{(1)}$ and $\cW^{(2)}$. (d) Composite domain wall $\cW^{(2)}\cW^{(1)}$.
}\label{fig:stable}
\end{figure}

\add{
The fusion spaces with four anyons further provide us consistence conditions of $\cW$.
To see this, first notice that there are \emph{generalised} tunneling channels,
$[O_{W,ia^*,x}]$,
which, in addition to tunneling $a$ to $i$, also create quasiparticle $x$ on the domain wall.
If we combine the tunneling channels $[O_{W,ia^*,x}]$ and $[O_{W,jb^*,x^*}]$,
we can create fusion states with a domain wall $W$ and four anyons $i,j,a^*,b^*$,  as Fig.~\ref{fig:stable}(a).
In other words, $[O_{W,ia^*,x}O_{W,jb^*,x^*}]$ form a basis of the fusion space $\mathcal V(S^2,i,j,W,a^*,b^*)$.
Let $\cK_{ia}^x$ denote the number of tunneling channels $[O_{W,ia^*,x}]$,
and we know that $\dim\mathcal V(S^2,i,j,W,a^*,b^*)=\sum_x \cK_{ia}^x \cK_{jb}^{x^*}$.
However, the tunneling process as Fig.~\ref{fig:stable}(b),
i.e., fusing $a,b$ to $c$, using $[O_{W,kc^*}]$ to tunnel $c$ to $k$ and splitting $k$ to $i,j$, forms another basis of the fusion space.
The number of such fusion/tunneling/splitting channels is $\sum_{kc} (\cN^B)_{ij}^k \cW_{kc} (\cN^A)_{ab}^c$.
Therefore, we must have 
\begin{equation}
  \sum_x \cK_{ia}^x \cK_{jb}^{x^*}= \sum_{kc} (\cN^B)_{ij}^k \cW_{kc} (\cN^A)_{ab}^c.
  \label{estable}
\end{equation}
}

We are interested in classifying \emph{stable} gapped domain walls,
i.e., the GSD cannot be
reduced no matter what small perturbations are added near the domain wall.
\add{
For stable gapped domain walls we have $\cW_{ia}=\cK_{ia}^{1}$.}
Unstable gapped domain walls $\mathcal{U}$ split as the sum of stable ones $\cW^{(1)},\cW^{(2)},\dots,\cW^{(N)}$, and $ \mathcal{U}_{ia}=\sum_{n=1}^N\cW^{(n)}_{ia}$, for $N\geq 2$.

\del{We find that a
gapped domain wall is stable \emph{if and only if} (iff) the tunneling matrix $\cW$ satisfies
the \emph{stable condition} \eqref{stable}:
$\cW_{ia}\cW_{jb}\leq\sum_{kc} (\cN^B)_{ij}^k \cW_{kc} (\cN^A)_{ab}^c$.
It can be understood in the following way.
Consider the number of channels tunneling $a,b$ to $i,j$ through the domain wall.
We may tunnel $a$ to $i$ and
$b$ to $j$ separately. The number of channels is $\cW_{ia}\cW_{jb}$. 
But this way,
we may miss some nontrivial exchanging channels $x$ along the domain wall as
Fig.~\ref{fig:stable}(a). 
If we first fuse
$a,b$ to $c$, tunnel $c$ to $k$ and then split $k$ to $i,j$,
instead we will obtain the
total number of channels $\sum_{kc} (\cN^B)_{ij}^k \cW_{kc} (\cN^A)_{ab}^c$,
as Fig.~\ref{fig:stable}(b).
The missing of exchanging channels leads to the inequality \eqref{stable}.
Such channel counting works only when the gapped domain wall is stable, so
\eqref{stable} is a necessary condition.
(This will be discussed further in Supplemental Material.)
But \eqref{commute}\eqref{stable}
together imply that $\cW_{11}=1$, thus $\cW$ cannot be the sum of more than one stable tunneling matrix; it must be stable itself.
Therefore, \eqref{stable} with \eqref{commute} is also sufficient
for a gapped domain wall to be stable.}
\add{Now, if a gapped domain wall $\cW$ is stable, \eqref{estable} becomes $  \sum_{kc} (\cN^B)_{ij}^k \cW_{kc} (\cN^A)_{ab}^c=\cW_{ia}\cW_{jb}+\sum_{x\neq 1} \cK_{ia}^x \cK_{jb}^{x^*}\geq \cW_{ia}\cW_{jb}$.
We know that \eqref{stable} is necessary for a gapped domain wall to be stable.
Furthermore, setting $i=j=a=b=1$ we know that $\cW_{11}\geq \cW_{11}^2$ and \eqref{commute} requires that $\cW_{11}>0$, thus $\cW_{11}=1$ and $\cW$ cannot be the sum of more than one stable tunneling matrix; it must be stable itself.
Therefore, \eqref{stable} with \eqref{commute} is also sufficient
for a gapped domain wall to be stable.
}

\se{Stability of composite domain walls}--
Let us consider two stable domain walls, $\cW^{(1)}$ between Phases $A$ and $B$, and
$\cW^{(2)}$ between Phases $B$ and $C$, as in Fig.~\ref{fig:stable}(c).
When the two domain walls are far separated, they are both stable.
Any small perturbations added near $\cW^{(1)}$, or near $\cW^{(2)}$, cannot reduce the GSD.

We then shrink the size of the middle Phase $B$, such that the two
domain walls are near enough to be regarded as a single domain wall.
This way we obtain a composite domain wall, whose tunneling matrix is the
composition $\cW^{(2)}\cW^{(1)}$, as Fig.~\ref{fig:stable}(d).
However, this composite domain wall $\cW^{(2)}\cW^{(1)}$ may no longer be stable.
\add{Unless Phase $B$ is vacuum,} we allow more perturbations to $\cW^{(2)}\cW^{(1)}$ than when 
$\cW^{(1)}$ and $\cW^{(2)}$ are far separated. Some operators simultaneously
acting on both $\cW^{(1)}$ and $\cW^{(2)}$ may reduce the GSD,
in which case, the composite domain wall $\cW^{(2)}\cW^{(1)}$ is not stable.

\add{In the special case when Phase $B$ is vacuum, the composite $\cW^{(2)}\cW^{(1)}$ remains stable.}
One can explicitly check this with \eqref{stable}. 

\se{GSD in the presence of gapped domain walls}--
\begin{figure}
  \centering
  \includegraphics[scale=1]{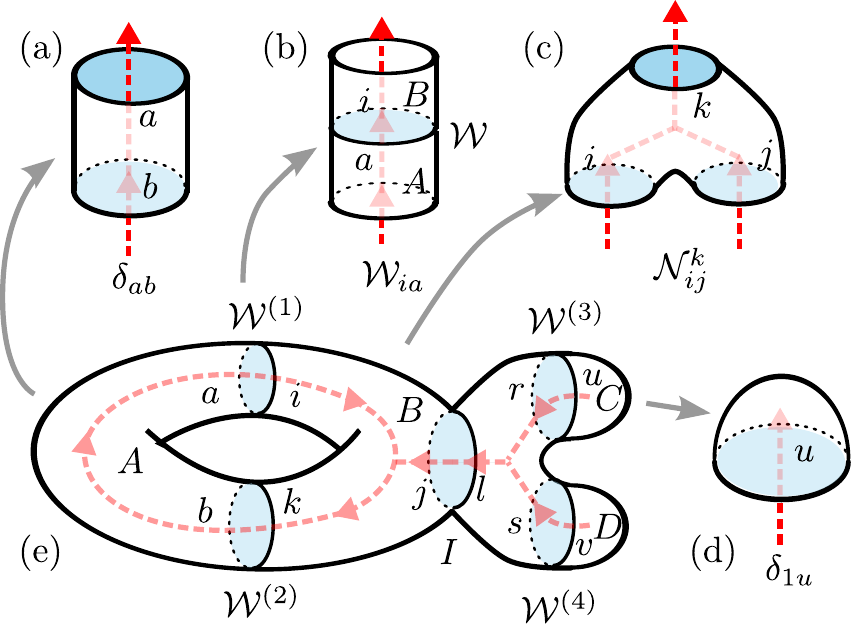}
  \caption{Computing GSD by tensor contraction: Cut a complicated manifold (e)
    into simple segments, add oriented skeletons and anyon indices.
    Associate the segments
    with:  (a) a cylinder with $\delta_{ab}$, 
    (b) a domain wall with its tunneling matrix $\cW_{ia}$, 
    (c) a pair of pants with the fusion tensor $\cN_{ij}^k$
    and (d) a cap with $\delta_{1u}$. Finally, contract all the tensors.
  }
  \label{fig:gdwfigure2}
\end{figure}
Below we derive the GSD, for a 2D system containing several
topological orders separated by loop-like gapped domain walls. 
Domain walls cut a whole 2D system into several segments. 
Without losing generality, let us consider an example in Fig.~\ref{fig:gdwfigure2} 
with topological orders, Phases $A,B,C,D$, and four nontrivial domain walls,
$\cW^{(1)},\cW^{(2)},\cW^{(3)},\cW^{(4)}$, on a manifold Fig.~\ref{fig:gdwfigure2}(e).
We first add extra trivial domain walls $\cW=\iden$, so that all segments
between domain walls are reduced to simpler topologies: caps, cylinders or
pants.
We also add oriented skeletons to the manifold, and put anyon indices
on both sides of the domain walls, 
shown in Fig.~\ref{fig:gdwfigure2}(e).
Next, see Fig.~\ref{fig:gdwfigure2}(a)(b)(c)(d), for the segments with oriented skeletons and anyon indices,
we associate certain tensors: caps with $\delta_{1u}$, cylinders with
$\delta_{ab}$, pants with $\cN_{ij}^k$ in the corresponding topological order,
and domain walls with their tunneling matrices $\cW_{ia}$. 
We may reverse the orientation and at the same time replace the
index $a$ with $a^*$.
Finally, we multiply these tensors together and contract all the anyon
indices. 
Physically, such tensor contraction computes the total number of winding channels
of anyons, which exactly counts the number of ground states, thus the GSD.

Systems with \emph{gapped boundaries} are
included in our method; just imagine that there are vacuum on
caps connected to the boundaries, e.g., Phases $C,D$ in Fig.~\ref{fig:gdwfigure2}(e)
can be vacuum. Dimensions of generic fusion spaces can also be
calculated, by putting the anyon $a$ on the cap and associating the tensor
$\delta_{au}$ instead of $\delta_{1u}$.

We derive GSD
on exemplary manifolds: 
\begin{enumerate}
  \item A stable domain wall $\cW$ on the sphere: $\GSD=\cW_{11}=1$.
  
  \item A domain wall $\cW$ on the torus: $\GSD=\Tr(\cW)$. 
  Several domain walls $\cW^{(1)},\dots,\cW^{(n)}$ on the torus,
  in Fig.~\ref{fig:gdwfigure}(a): $\GSD=\Tr(\cW^{(1)}\cdots \cW^{(n)})$.
  In particular, $\Tr[\cW^{(1)}(\cW^{(2)})^\dag]$ counts the types of \emph{0D defects} between 1D gapped domain walls $\cW^{(1)},\cW^{(2)}$.
  
%

  \item A sphere with punctures: A cylinder with two gapped boundaries $\cW^L$ and $\cW^R$, in Fig.~\ref{fig:gdwfigure}(b): $\GSD=\sum_a \cW^L_{a1} \cW^R_{1a}$.
  A pair of pants with three gapped boundaries $\cW^{(1)}$, $\cW^{(2)}$ and $\cW^{(3)}$, in Fig.~\ref{fig:gdwfigure}(c): $\GSD={\sum}_{ijk} \cW^{(1)}_{i1} \cW^{(2)}_{j1} \cN^k_{ij} \cW^{(3)}_{1k}$.
  
  \item The rocket graph in Fig.~\ref{fig:gdwfigure2}(e): $\GSD
  =\underset{{a,i,j,k,r,s}}{\sum} \cW^{(1)}_{ia} \cW^{(2)}_{ak}  (\cN^{B})^k_{ij} (\cN^{B})^j_{rs} \cW^{(3)}_{r1} \cW^{(4)}_{s1}$.
\end{enumerate}
We apply our formalism to several topological orders.
Details of our examples are organized in Supplemental Material.
Part of our result is listed in Table \ref{table:gappedDW} (the number of gapped domain walls types) and Table \ref{table:GSDpunc} (GSD).

\begin{table}[!h]
\raggedright
  \begin{minipage}[b]{0.45\linewidth}
    {\begin{tabular}{ |c|   c | c|}
  \hline
\# gapped DW  &  vacuum &  toric code  \\
  \hline
toric code &   2 & 6 \\
double-semion &  1 & 2\\ 
doubled Fibonacci & 1  & 2 \\ 
doubled Ising &  1 &  3\\ 
$D(S_3)$ & 4  &  12\\ 
  \hline
\end{tabular} }
  \end{minipage}\;\;\;\;\;\;\;\;\;\;\;\;\;\;\;\;\;\;
\begin{minipage}[b]{0.35\linewidth}
{\begin{tabular}{ |c|   c | }
  \hline
\# gapped DW  &  vacuum    \\
  \hline
 $D(D_4)$ &   11 \\
$D(Q_8)$ &  6 \\ 
$D^{\omega_3{[3d]}}(Z_2{}^3)$ & 5 \\ 
$D^{\omega_3{[5]}}(Z_2{}^3)$ &  3\\ 
$D^{\omega_3{[7]}}(Z_2{}^3)$ & 1 \\ 
  \hline
\end{tabular} }
  \end{minipage}
\caption{The number of different gapped domain wall types (``\# gapped DW'' for short) sandwiched by two topological orders (one from the first column and the other from the first row).
$D^{\omega_3}(G)$ stands for the twisted quantum double model of gauge group $G$ with a 3-cocycle twist ${\omega_3}$.}  
\label{table:gappedDW}
\end{table}
\begin{table}[!h]
\centering
\begin{tabular}{ |c|   c|  c |  c| c| }
  \hline
GSD(\# punctures) &\;1\;&  2 & 3 & 4   \\
  \hline
 toric code  & \,1\,& 1, 2 &  2, 4 & 2, 4, 8  \\
double-semion &\,1\,&  2 &  4 & 8  \\
doubled Fibonacci&\,1\, & 2 & 5  & 15  \\
doubled Ising & \,1\,& 3 & 10 & 36\\
  \hline
\end{tabular}
\caption{ GSD of a single topological order (the first column) on a sphere with a number of punctures (the first row).
Each puncture has a gapped boundary. 
The last three orders allow only one type of gapped boundary, so its GSD is unique for a given topology.
Toric code allows two types of gapped boundaries, and its GSD varies, which depends on boundary types associated to each puncture.
This agrees with~\cite{WW12,BK98}.}  
\label{table:GSDpunc}
\end{table}

\se{Conclusion}-- Given $\cS,\cT$ matrices of topological orders with the
same central charge, we have provided simple criteria \eqref{Winteger}\eqref{commute}\eqref{stable} to check the existence of gapped domain walls. 
We want to mention that,
a gapped domain wall can be related to a gapped boundary
by the \emph{folding trick}~\cite{KK12}.
By studying gapped boundaries, we can also obtain all the information of gapped domain walls.
But, to compute the GSD, gapped domain walls allow more configurations
on 2D surfaces than gapped boundaries.

The gapped domain walls and boundaries can be explicitly realized in lattice
models~\cite{BSW11,KK12,LW14}.
Levin-Wen string-net models~\cite{LW05} are exactly solvable models for
topological orders. Recently it was found that a topological order can be
realized by a Levin-Wen model iff 
it has gapped
boundaries~\cite{KK12,LW14}.
Thus, our work provides the criteria whether a topological order has a
Levin-Wen realization.

2D Abelian topological orders can be
described by Chern-Simons field theories. The boundary of a Chern-Simons
theory is gappable, iff there exists a \emph{Lagrangian subgroup}~\cite{KS11,WW12,Lev13,BJQ13,BJQ13a,Kap14}.
Our tunneling matrix criteria \eqref{Winteger}\eqref{commute}\eqref{stable}
are equivalent to the Lagrangian subgroup criteria
for Abelian topological orders (a detailed proof is given in Supplemental Material), but are more general and also apply to non-Abelian topological orders.

One can also use the \emph{anyon condensation} approach~\cite{BS09,BSH09,FSV13,Kon14,ERB13,HW1308,HW1408} to determine the gapped boundaries
of (non-Abelian) topological orders,
by searching for the \emph{Lagrangian condensable anyons}
(mathematically, \emph{Lagrangian algebras}~\cite{{FSV13},{Kon14}}),
whose condensation will break the topological order to vacuum.
However, we use only an integer vector $\cW_{1a}$
to determine the anyon $\bm q_\cW$,
while in the anyon condensation approach,
besides the multiplicity $\cW_{1a}$, there are many additional data 
satisfying a series of formulas.
These formulas put certain constraints on the condensable anyon,
but not in a simple and explicit manner.
Our claim that \eqref{Winteger}\eqref{commute}\eqref{stable} are necessary and sufficient for a gapped domain wall to exist means that,
Lagrangian condensable anyons must satisfy \eqref{Winteger}\eqref{commute}\eqref{stable}, and, 
for the anyon $\bm q_\cW$ satisfying \eqref{Winteger}\eqref{commute}\eqref{stable},
there must exist solutions to the additional data in the anyon condensation approach.

We know that the effective 1+1D edge theory of a 2+1D topological order has
a gravitational anomaly. The gravitational anomalies are classified by the bulk topological order
$(\cS,\cT,c_-)$~\cite{W1313,KW14}.  When $c_-\neq 0$, the edge effective theory
has a perturbative  gravitational anomaly which leads to topological gapless
edge (i.e., the gaplessness of the edge is robust against any change of
the edge Hamiltonian).  Even in the absence of perturbative  gravitational
anomaly, $c_-=0$, certain global gravitational anomalies~\cite{Witten:1985xe} (characterized by
$(\cS,\cT,0)$) can also lead to topological gapless edge~\cite{WW12, Lev13}.  Our
work points out that such global gravitational anomalies are described by
$\cS,\cT$ which do not allow any non-zero solution $\cW$ of
\eqref{Winteger}\eqref{commute}\eqref{stable}. The corresponding 2D
topological order $(\cS,\cT,0)$ will have topological gapless edge.

Since a domain wall sits on the border between two topological orders, 
our study on domain walls can also guide us to better understand the \emph{phase transitions} of topological orders.

%

\begin{acknowledgments}
\se{Acknowledgements}--
After posting the arXiv preprint version 1, we 
are grateful receiving very helpful comments from Liang Kong, John Preskill, Anton Kapustin and Yidun Wan.
This research is supported by NSF Grant No.\;DMR-1005541, NSFC 11074140, NSFC
11274192,
the BMO Financial Group and
the John Templeton Foundation. Research at Perimeter Institute is supported by the Government of Canada
through Industry Canada and by the Province of Ontario
through the Ministry of Research.

\se{Note added}--
During the preparation of this manuscript, we become aware that a recent work
Ref.~\cite{HW1408} has independently obtained part of our results using a different approach: anyon condensation. 
The comparison between our new approach and anyon condensation is explained in Conclusion.
\end{acknowledgments}
\bibliography{gdwref}

\newpage
\appendix
{\centering\large\bf Supplemental Material\par}

\del{\section{Sec. 1. The stable condition and quasiparticles on the gapped domain wall}
Here we clarify the relation between the stability of gapped domain walls
and the quasiparticles on them.

The exchanging channel $x$,
mentioned 
in the main text,
actually corresponds to the quasiparticle $x$ on the domain wall $\cW$
between Phases $A$ and $B$.
Similar to the bulk anyons, domain wall quasiparticles can also fuse;
we use another fusion tensor to describe their fusion
$x\ot y=\oplus_z \cK_{xy}^z z$.
But they can not braid on a 1D domain wall (we do not call them anyons).
To fully describe the tunneling, we need another tensor $\cK_{ia}^x$,
which means that, 
if we tunnel an anyon $a$ in Phase $A$,
it becomes $\oplus_{ix} \cK_{ia}^x (i\ot x)$,
where $i$ is in Phase $B$ while $x$ is stuck on the domain wall.
$\cK_{ia}^x$ must be compatible with the fusion tensors
$(\cN^A)_{ab}^c$, $(\cN^B)_{ij}^k$ and $\cK_{xy}^z$, i.e.,
\[ 
  \sum_{xy} \cK_{xy}^z \cK_{ia}^x\cK_{jb}^y
=\sum_{kc} (\cN^B)_{ij}^k \cK_{kc}^z (\cN^A)_{ab}^c\,.\]

We denote the trivial quasiparticle on the domain wall $\cW$
(namely the ground state of $\cW$) by $\one_\cW=\oplus_x \one_\cW^x x$. 
The domain wall is stable iff $\one_\cW$ contains only one quasiparticle label,
i.e., $\one_\cW^x=\delta_{1x}$.
The tunneling matrix $\cW$ counts the tunneling channels
where the quasiparticles stuck on the domain wall is the trivial one $\one_\cW$.
That is to say, $\cW_{ia}=\sum_x\one_\cW^x \cK_{ia}^x$. Thus,
\[\sum_{kc} (\cN^B)_{ij}^k \cW_{kc} (\cN^A)_{ab}^c=
\sum_{xyz} \one_\cW^z \cK_{xy}^z \cK_{ia}^x\cK_{jb}^y.\]
If $\cW$ is stable, $\one_\cW^z=\delta_{1z}$, $\cW_{ia}=\cK_{ia}^1$, we have
\begin{align*}
  \sum_{kc} (\cN^B)_{ij}^k \cW_{kc} (\cN^A)_{ab}^c
  &=\sum_{xy}\cK_{xy}^1 \cK_{ia}^x\cK_{jb}^y\\
  &=\cK_{ia}^1\cK_{jb}^1
+\sum_{x\neq 1\text{ or }y\neq 1}\cK_{xy}^1 \cK_{ia}^x\cK_{jb}^y\\
&=\cW_{ia}\cW_{jb}+\sum_{x\neq 1} \cK_{ia}^x\cK_{jb}^{x^*}\\
&\geq \cW_{ia}\cW_{jb},
\end{align*}
which is exactly the stable condition.
The term $\sum_{x\neq 1} \cK_{ia}^x\cK_{jb}^{x^*}$ counts the number
of nontrivial exchanging channels.}

\section{Sec. 1.  Equivalence between tunneling matrices and Lagrangian subgroups for Abelian topological orders}
If the fusion of anyons has a group structure, $N_{ab}^c=\delta_{a+b,c}$,
the topological order is called \emph{Abelian}.
We denote the fusion group by $\cL$. Since $\cL$ is an Abelian group,
we label the trivial type by $0$ instead of $1$.
The rank of $\cT,\cS$ matrices is $|\cL|$, and
$ \forall a$, $\cS_{0a}={|\cL|}^{-1/2}$.
We rescale the $\cS$ matrix as
\begin{align*}
  \tS=\sqrt{|\cL|}\cS.
\end{align*}
The Verlinde formula then implies
\begin{align*}
  \tS_{ac}\tS_{bc}=\tS_{a+b,c},
\end{align*}
which means that for each $c$, $\tS_{-,c}$ form a 1D linear representation of the
fusion group $\cL$. $\tS$ matrix is the character table of $\cL$.

The gapped boundaries of Abelian topological orders
are classified by \emph{Lagrangian subgroups}.
We introduce a physical definition of Lagrangian subgroup,
in terms of anyon statistics ($\cT,\cS$ matrices).~\cite{Lev13}
A Lagrangian subgroup $\cM$ is a subset of anyons, $\cM\in\cL$, such that\\ 
(i) If $a\in \cM$, then $\cT_{aa}=\re^{\ii \theta_a}=1$.\\
(ii) If $a \in \cM$, then $\forall c\in \cM,$ $ \tS_{ac}=1$.\\
(iii) If $a \not\in \cM$, then $\exists c\in \cM,$ $ \tS_{ac}\neq 1$.\\
(iii) is equivalent to\\
(iii)' If $\forall c\in \cM,$ $ \tS_{ac}=1$, then $a\in \cM$.\\
Note that (ii) and (iii)' implies that $\cM$ is a subgroup.
First, $\forall c\in \cM,$ $ \tS_{0c}=1$, thus, the identity $0$ is in $\cM$.
Second, if $a\in\cM$ and $b\in\cM$, then $\forall c\in\cM,$ $
\tS_{ac}=\tS_{bc}=1$.
Thus, $\forall c\in\cM,$ $\tS_{a+b,c}=\tS_{ac}\tS_{bc}=1$ and $a+b\in\cM$.
Finally, $\tS_{-a,c}=\tS_{ac}^{-1}$ so if $a\in\cM$ we also have $-a\in\cM$.

Next, we want to show that,
for gapped boundaries of Abelian topological orders,
Lagrangian subgroups are in one-to-one correspondence
with tunneling matrices.
First consider the stable condition. 
It reduces to $\cW_{a+b}\geq\cW_{a}\cW_{b}$.
(We omit the anyon index of the vacuum.)
In particular, $1=\cW_{0}\geq\cW_{a}\cW_{-a}=\cW^2_{a}$,
which implies that $\cW_{a}\leq 1$.
Thus, we relate a Lagrangian subgroup $\cM$ and a tunneling matrix $\cW$ via
\begin{align*}
  a\in\cM &\Leftrightarrow \cW_{a}=1,\\
  a\not\in\cM&\Leftrightarrow \cW_{a}=0.
\end{align*}

It is easy to see (i) is equivalent to $\cT \cW=\cW$.
We will focus on the proof of (ii)(iii)~$\Leftrightarrow \cS\cW=\cW$.

(ii)(iii)~$\Leftarrow \cS\cW=\cW$ is easier.
Consider the first row
\[\sum_{a}\cS_{0a}\cW_{a}=\sum_{a\in\cM}\cS_{0a}
=\frac{|\cM|}{\sqrt{|\cL|}}=\cW_0=1.\]
We have $|\cM|=\sum_a \cW_a=\sqrt{|\cL|}$ and
\[\cW_a=\sum_c \cS_{ac}\cW_c= \sum_{c}\frac{\tS_{ac}}{\sqrt{|\cL|}}\cW_c
=\sum_{c\in\cM}\frac{\tS_{ac}}{|\cM|}.\]
Now, if $a\in\cM$, i.e., $\cW_a=1$,
$\tS_{ac}$ in the above equation must all be 1.
(Note that $\tS_{ac}$ are all phase factors $|\tS_{ac}|=1$).
This is (ii).
If $a\not\in\cM$, i.e., $\cW_a=0$, there must be at least one $\tS_{ac}\neq 1$
in the above equation. This is (iii).

The other direction (ii)(iii)~$\Rightarrow \cS\cW=\cW$ is a bit involving.
First, note the following relation
\begin{align*}
  \tS_{ac}=\tS_{bc},\forall c\in\cM
  &\Leftrightarrow \tS_{a-b,c}=1, \forall c\in\cM\\
  &\Leftrightarrow a-b \in \cM \Leftrightarrow a\in b+\cM.
\end{align*}
This motivates us to consider the quotient group $\cL/\cM$.
Each element $b+\cM$ in $\cL/\cM$ gives rise to a 1D representation of $\cM$,
i.e., $\tS_{b,-}$.
Different elements in $\cL/\cM$ gives different 1D representations.
Since the Abelian group $\cM$ has in total $|\cM|$ different 1D representations,
we have $|\cL/\cM|=|\cL|/|\cM|\leq |\cM|$.
On the other hand, each $c\in\cM$ gives rise to a 1D representation of $\cL/\cM$,
i.e., $\tS_{-,c}$.
Since the $\cS$ matrix is invertible, different $c\in\cM$ gives
different 1D representations of $\cL/\cM$.
Again, $\cL/\cM$ has in total $|\cL|/|\cM|$ different 1D representations.
We also have $|\cM|\leq |\cL|/|\cM|$.
Thus, we know that $|\cM|=|\cL|/|\cM|$, i.e., $|\cM|=\sqrt{|\cL|}$.

Now, if $a\in\cM$, we have
\[\sum_{c} \cS_{ac}\cW_c= \sum_{c}\frac{\tS_{ac}}{\sqrt{|\cL|}}\cW_c
=\sum_{c\in\cM} \frac{1}{|\cM|}=1=\cW_a.\]
If $a\not\in\cM$, then $\exists c\in\cM,$ $ \tS_{ac}\neq 1$.
In other words, $\tS_{a,-}$ is a nontrivial 1D representation of $\cM$,
and we know that $\sum_{c\in\cM} \tS_{ac}=0$. Thus,
\[\sum_{c}\cS_{ac}\cW_c=\frac{1}{|\cM|}\sum_{c\in\cM}\tS_{ac}=0=\cW_a.\]
We have proved that (ii)(iii)~$\Rightarrow \cS\cW=\cW$.

To conclude, for gapped boundaries
(and gapped domain walls by the folding trick) of Abelian topological orders,
our tunneling matrix criteria 
is equivalent to the Lagrangian subgroup criteria.

\section{Sec. 2.  Examples}
We provide explicit data of gapped boundaries and gapped domain walls of 2D topological orders,
computed by our formalism developed 
in the main text.  

A list of topological orders we consider contains (with their notations of twisted quantum double model $D^{\omega_3[n]}(G)$ 
for a gauge group $G$ with a 3-cocycle twist $\omega_3$, and $n$ implies the number of pairs of $\pm \ii$ in its $\cT$ matrix.):\\

\noindent
(i). toric code ($D(\mathbb{Z}_2)$),\\
(ii). double-semion ($D^{\omega_3[1]}(\mathbb{Z}_2)$),\\
(iii). doubled Fibonacci phase (Fibonacci $\times$ $\overline{\text{Fibonacci}}$ ),\\
(iv). doubled Ising phase (Ising $\times$ $\overline{\text{Ising}}$ ),\\
(v). $D(S_3)$ as the quantum doubled model of the permutation group $S_3$ of order 6,\\
(vi). $D(D_4)=D^{\omega_3{[1]}}(Z_2{}^3)$ as the quantum doubled model of the dihedral group $D_4$ of order 8,\\
(vii). $D(Q_8)=D^{\omega_3{[3i]}}(Z_2{}^3)$ as the quantum doubled model of the quaternion group $Q_8$ of order 8,\\
(viii). $D^{\omega_3{[3d]}}(Z_2{}^3)$ as a twisted quantum doubled model of the group $\mathbb{Z}_2{}^3$ of order 8 with a 3-cocylce twist ${\omega_3{[3d]}}$,\\
(ix). $D^{\omega_3{[5]}}(Z_2{}^3)$ as a twisted quantum doubled model of the group $\mathbb{Z}_2{}^3$ of order 8  with a 3-cocylce twist ${\omega_3{[5]}}$,\\
(x). $D^{\omega_3{[7]}}(Z_2{}^3)$ as a twisted quantum doubled model of the group $\mathbb{Z}_2{}^3$ of order 8  with a 3-cocylce twist ${\omega_3{[7]}}$.\\

One may refer to Refs.~\cite{Propitius95,HWW11} for an introduction to twisted quantum double models.

Here ${\omega_3{[3i]}}$ means a 3-cocycle whose $D^{\omega_3{[3i]}}$ model generates 3 pairs of $\pm \ii$ in its $\cT$ matrix and their generators are linear independent ($i$).
${\omega_3{[3d]}}$ means a 3-cocycle whose $D^{\omega_3{[3d]}}$ model generates 3 pairs of $\pm \ii$ in its $\cT$ matrix and their generators are linear dependent ($d$).
More detail are explained in Ref.~\cite{WangWen1404} and reference therein.

Below we will provide $\cS,\cT$ matrices, 
tunneling matrices $\cW$ of gapped boundaries and gapped domain walls of these topological orders (i)-(x).
We will count \emph{the number of types} of gapped boundaries and gapped domain walls.
We will also count some examples of their \emph{ground state degeneracy} (GSD) on various manifolds with gapped boundaries on the punctures.

The $\cS,\cT$ matrices of all five kinds of non-Abelian twisted quantum double models $D^{\omega_3{}}(Z_2{}^3)$ are explicitly adopted from 
the calculation of Ref.~\cite{WangWen1404}. 
%

\subsection{{I. Gapped boundaries of toric code phase: 2 types}}

The $\cS,\cT$ matrices of toric code phase are:
\begin{align*}
  \cT&=\Diag(1,1,1,-1),\\
  \cS&=\frac{1}{2}\begin{pmatrix}
    1&1&1&1\\
    1&1&-1&-1\\
    1&-1&1&-1\\
    1&-1&-1&1
  \end{pmatrix}.
\end{align*}
There are two types of gapped boundaries:
\begin{align*}
  \cW^\text{TC}_e&=\begin{pmatrix} 1&1&0&0 \end{pmatrix},\\
  \cW^\text{TC}_m&=\begin{pmatrix} 1&0&1&0 \end{pmatrix}.
\end{align*}
Conventionally, we label the 4 types of anyons (quasiparticles) as $1,e,m,\varepsilon$.
The $\cW^\text{TC}_e$ boundary corresponds to condensing $e$ and $\cW^\text{TC}_m$ corresponds
to condensing $m$.

   We compute GSD on a cylinder with two gapped boundaries.
   Note that the GSD is also the number of types of 0D defects between the two gapped boundaries. In particular, it is the number of boundary quasiparticle types if the two gapped boundaries are the same. 
\begin{align*}
  \begin{array}{c|cc}
    \GSD &(W^{TCe})^\dag&(W^{TCm})^\dag\\
    \hline
    W^{TCe}&2&1\\
    W^{TCm}&1&2
  \end{array}
\end{align*}
This agrees with~\cite{WW12,BK98}.

\subsection{II. Gapped domain walls between two toric codes: 6 types}

There are 6 types of gapped domain walls between two toric codes. The first two are invertible (transparent domain walls):
\begin{align*}
  \cW^\mathrm{TC|TC}&=I,\\
  \cW^\mathrm{TC|TC}_{e\leftrightarrow m}&=\begin{pmatrix}
    1&0&0&0\\
    0&0&1&0\\
    0&1&0&0\\
    0&0&0&1
  \end{pmatrix}.
\end{align*}
The rest 4 are the compositions of gapped boundaries,
i.e., $(\cW^\text{TC}_e)^\dag\cW^\text{TC}_e,$
$(\cW^\text{TC}_e)^\dag\cW^\text{TC}_m,$
$(\cW^\text{TC}_m)^\dag\cW^\text{TC}_e,$
$(\cW^\text{TC}_m)^\dag\cW^\text{TC}_m$.
The GSD on the torus with the
$e,m$-exchanging domain wall $\cW^\mathrm{TC|TC}_{e\leftrightarrow m}$~\cite{KK12,YouWen12} turns out to be
${\Tr(\cW^\mathrm{TC|TC}_{e\leftrightarrow m})=2}$.


\subsection{III. Gapped boundary of double-semion phase: 1 type}

The $\cS,\cT$ matrices of double-semion phase are:

\begin{align*}
  \cT&=\Diag(1,\ii,-\ii,1),\\
  \cS&=\frac{1}{2}
  \begin{pmatrix}
    1&1&1&1\\
    1&-1&1&-1\\
    1&1&-1&-1\\
    1&-1&-1&1
  \end{pmatrix}.
\end{align*}
There is only one type of gapped boundary,
\begin{align*}
  \cW^\text{DS}=\begin{pmatrix}
    1&0&0&1
  \end{pmatrix}.
\end{align*}
Since there is only one gapped boundary type, we would like to compute the GSD on spheres with
more punctures:
\begin{itemize}
  \item 2 punctures (a cylinder): GSD=2,
  \item 3 punctures (a pair of pants): GSD=4,
  \item 4 punctures: GSD=8,
  \item 5 punctures: GSD=16,
  \item $n$ punctures: GSD=$2^{n-1}$.
\end{itemize}


\subsection{IV. Gapped domain walls between double-semion and toric code phases: 2 types}

Gapped domain walls between double-semion and toric code phases only have two types. They are the compositions of gapped boundaries,
$(\cW^\text{TC}_{e})^\dag
\cW^\text{DS}$ and $(\cW^\text{TC}_{m})^\dag
\cW^\text{DS}$.


\subsection{V. Gapped boundary of doubled Fibonacci phase: 1 type}

Let
$\gamma=\dfrac{1+\sqrt{5}}{2}$. The $\cS,\cT$ matrices  of doubled Fibonacci phase are:
\begin{align*}
  \cT&=\Diag(1,\re^{-\frac{4\pi \ii}{5}},\re^{\frac{4\pi \ii}{5}},1),\\
  \cS&=\frac{1}{1+\gamma^2}
  \begin{pmatrix}
    1&\gamma&\gamma&\gamma^2\\
    \gamma &-1&\gamma^2&-\gamma\\
    \gamma&\gamma^2&-1&-\gamma\\
    \gamma^2&-\gamma&-\gamma&1
  \end{pmatrix}.
\end{align*}
There is only one type of gapped boundary,
\begin{align*}
  \cW^\text{DF}=\begin{pmatrix}
    1&0&0&1
  \end{pmatrix}.
\end{align*}
We compute the GSD on spheres with more punctures:
\begin{itemize}
  \item 2 punctures (cylinder): GSD=2,
  \item 3 punctures (a pair of pants): GSD=5,
  \item 4 punctures: GSD=15,
\end{itemize}


\subsection{VI. Gapped domain walls between doubled Fibonacci and toric code phases: 2 types}

Gapped domain walls between doubled Fibonacci and toric code phases only have two types. They are the compositions of gapped boundaries,
$(\cW^\text{TC}_{e})^\dag
\cW^\text{DF}$ and $(\cW^\text{TC}_{m})^\dag
\cW^\text{DF}$.


\subsection{VII. Gapped boundary of doubled Ising phase: 1 type}

Let $\varphi=\sqrt{2}$. The $\cS,\cT$ matrices of doubled Ising phase are
\begin{align*}
  \cT&=\Diag(1,\re^{-\frac{\pi \ii}{8}},-1,\re^{\frac{\pi
  \ii}{8}},1,-\re^{\frac{\pi \ii}{8}},-1,-\re^{\frac{-\pi \ii}{8}},1),\\
  \cS&=\frac{1}{4}
  \left(
  \begin{array}{ccccccccc}
    1 & \varphi & 1 & \varphi & 2 & \varphi & 1 & \varphi & 1 \\
    \varphi & 0 & -\varphi & 2 & 0 & -2 & \varphi & 0 & -\varphi \\
    1 & -\varphi & 1 & \varphi & -2 & \varphi & 1 & -\varphi & 1 \\
    \varphi & 2 & \varphi & 0 & 0 & 0 & -\varphi & -2 & -\varphi \\
    2 & 0 & -2 & 0 & 0 & 0 & -2 & 0 & 2 \\
    \varphi & -2 & \varphi & 0 & 0 & 0 & -\varphi & 2 & -\varphi \\
    1 & \varphi & 1 & -\varphi & -2 & -\varphi & 1 & \varphi & 1 \\
    \varphi & 0 & -\varphi & -2 & 0 & 2 & \varphi & 0 & -\varphi \\
    1 & -\varphi & 1 & -\varphi & 2 & -\varphi & 1 & -\varphi & 1 \\
  \end{array}
  \right).  
\end{align*}
There is only one type of gapped boundary,
\begin{align*}
  \cW^\text{DI}=\begin{pmatrix}
    1&0&0&0&1&0&0&0&1
  \end{pmatrix}.
\end{align*}
We compute the GSD on spheres with more punctures:
\begin{itemize}
  \item 2 punctures (cylinder): GSD=3,
  \item 3 punctures (a pair of pants): GSD=10,
  \item 4 punctures: GSD=36,
\end{itemize}


\subsection{VIII. Gapped domain walls between doubled Ising and toric code phases: 3 types}

There are 3 types of stable gapped domain walls between doubled Ising and toric code phases. The first one is
\begin{align*}
\cW^{\mathrm{TC}|\mathrm{DI}}=
\left(
\begin{array}{ccccccccc}
 1 & 0 & 0 & 0 & 0 & 0 & 0 & 0 & 1 \\
 0 & 0 & 0 & 0 & 1 & 0 & 0 & 0 & 0 \\
 0 & 0 & 0 & 0 & 1 & 0 & 0 & 0 & 0 \\
 0 & 0 & 1 & 0 & 0 & 0 & 1 & 0 & 0 \\
\end{array}
\right)
\end{align*}
If we label the anyons in the doubled Ising phase as
$1\ov{1},1\ov{\sigma},1\ov{\psi},\sigma\ov{1},\sigma\ov{\sigma},\sigma\ov{\psi},\psi\ov{1},\psi\ov{\sigma},\psi\ov{\psi}$,
this domain wall corresponds to the follow tunneling process
\begin{gather*}
  1\ov{1}\to 1,\quad \psi\ov{\psi} \to 1,\\
  1\ov{\psi}\to \varepsilon,\quad \psi\ov 1 \to \varepsilon,\\
  \sigma\ov\sigma\to e\oplus m.
\end{gather*}
This agrees with a recent result obtained by anyon condensation~\cite{HW1308}.

The other two types of gapped domain walls, again, are the compositions of gapped boundaries,
$(\cW^\text{TC}_{e})^\dag
\cW^\text{DI}$ and $(\cW^\text{TC}_{m})^\dag
\cW^\text{DI}$.

We also like to use this example
to illustrate the instability of composite domain walls.
Insert a strip of doubled Ising phase to the toric code phase,
together with gapped domain walls $(\cW^{\mathrm{TC}|\mathrm{DI}})^\dag$ and $\cW^{\mathrm{TC}|\mathrm{DI}}$.
We then shrink the doubled Ising phase strip and compose the two domain walls.
By straightforward calculation,
\[\cW^{\mathrm{TC}|\mathrm{DI}}(\cW^{\mathrm{TC}|\mathrm{DI}})^\dag
=\left(
\begin{array}{cccc}
 2 & 0 & 0 & 0 \\
 0 & 1 & 1 & 0 \\
 0 & 1 & 1 & 0 \\
 0 & 0 & 0 & 2 \\
\end{array}
\right)
=I+\cW^\mathrm{TC|TC}_{e\leftrightarrow m}.\]
Thus, the composite domain wall splits to one trivial domain wall $I$,
and, one $e,m$ exchanging domain wall $\cW^\mathrm{TC|TC}_{e\leftrightarrow m}$,
between two toric code phases.
It is an unstable gapped domain wall
that does not satisfy the stable condition. 


\subsection{IX. Gapped boundaries of $D(S_3)$ phase: 4 types}

The $\cS,\cT$ matrices of $D(S_3)$ phase are
\begin{align*}
  \cT&=\Diag(1,1,1,1,\re^{-\frac{2\pi\ii}{3}},\re^{\frac{2\pi\ii}{3}},1,-1),\\
  \cS&=\frac{1}{6}
  \begin{pmatrix}
    1&1&2&2&2&2&3&3\\
    1&2&2&2&2&2&-3&-3\\
    2&2&4&-2&-2&-2&0&0\\
    2&2&-2&4&-2&-2&0&0\\
    2&2&-2&-2&4&-2&0&0\\
    2&2&-2&-2&-2&4&0&0\\
    3&-3&0&0&0&0&3&-3\\
    3&-3&0&0&0&0&-3&3
  \end{pmatrix}.
\end{align*}
There are 4 types of gapped boundaries.
\begin{align*}
  \cW^{D(S_3)}_{(1)}&= \begin{pmatrix} 1&1&2&0&0&0&0&0 \end{pmatrix} ,\\
  \cW^{D(S_3)}_{(2)}&= \begin{pmatrix} 1&1&0&2&0&0&0&0 \end{pmatrix},\\
  \cW^{D(S_3)}_{(3)}&= \begin{pmatrix} 1&0&1&0&0&0&1&0 \end{pmatrix} ,\\
  \cW^{D(S_3)}_{(4)}&=\begin{pmatrix} 1&0&0&1&0&0&1&0 \end{pmatrix} .
\end{align*}
We compute the GSD on a cylinder with two gapped boundaries (read from the above):
\begin{align*}
  \begin{array}{c|cccc}
    \GSD & (1)&(2)&(3)&(4)\\\hline
    (1) &6&2&3&1\\
    (2) &2&6&1&3\\
    (3) &3&1&3&2\\
    (4) &1&3&2&3
  \end{array}
\end{align*}


\subsection{X. Gapped domain walls between $D(S_3)$ and toric code: 12 types}

Gapped domain walls between $D(S_3)$ and toric code have 12 types in total. The first two types:
\begin{align*}
  \cW^{\mathrm{TC}|D(S_3)}_{(1)}&=\left(
\begin{array}{cccccccc}
 1 & 0 & 1 & 0 & 0 & 0 & 0 & 0 \\
 0 & 1 & 1 & 0 & 0 & 0 & 0 & 0 \\
 0 & 0 & 0 & 0 & 0 & 0 & 1 & 0 \\
 0 & 0 & 0 & 0 & 0 & 0 & 0 & 1 \\
\end{array}
\right),\\
  \cW^{\mathrm{TC}|D(S_3)}_{(2)}&=\left(
\begin{array}{cccccccc}
 1 & 0 & 0 & 1 & 0 & 0 & 0 & 0 \\
 0 & 1 & 0 & 1 & 0 & 0 & 0 & 0 \\
 0 & 0 & 0 & 0 & 0 & 0 & 1 & 0 \\
 0 & 0 & 0 & 0 & 0 & 0 & 0 & 1 \\
\end{array}
\right).
\end{align*}
The third and fourth types are the first two types composed with the $e,m$-exchanging
domain wall $\cW^\mathrm{TC|TC}_{e\leftrightarrow m}$,
i.e., $\cW^\mathrm{TC|TC}_{e\leftrightarrow
m}\cW^{\mathrm{TC}|D(S_3)}_{(1)}, \cW^\mathrm{TC|TC}_{e\leftrightarrow
m}\cW^{\mathrm{TC}|D(S_3)}_{(2)}$. The other 8 types are the compositions of gapped
boundaries (toric code has 2 types and $D(S_3)$ has 4 types of gapped boundaries ).



\newpage
\onecolumngrid
\subsection{XI. Gapped boundaries of $D(D_4)$ phase: 11 types}

Note that $D(D_4)=D^{\omega_3{[1]}}(Z_2{}^3)$.
 To simplify notations, below we denote
    $q_i=\begin{pmatrix}
      0&\cdots&0&1&0&\cdots&0
    \end{pmatrix}$ where 1 is the $i$th entry.
The $\cS,\cT$ matrices of $D(D_4)$ are~\cite{Propitius95,WangWen1404}:

\begin{align*}
  \cT&= \Diag(1,1,1,1,1,1,1,1,1,1,1,1,1,1,\ii,-1,-1,-1,-1,-1,-1,-\ii),\\
  \cS&=\frac{1}{8} \left(
  \begin{array}{cccccccccccccccccccccc}
    1 & 1 & 1 & 1 & 1 & 1 & 1 & 1 & 2 & 2 & 2 & 2 & 2 & 2 & 2 & 2 & 2 & 2 & 2 & 2 & 2 & 2 \\
    1 & 1 & 1 & 1 & 1 & 1 & 1 & 1 & -2 & 2 & 2 & -2 & -2 & 2 & -2 & -2 & 2 & 2 & -2 & -2 & 2 & -2 \\
    1 & 1 & 1 & 1 & 1 & 1 & 1 & 1 & 2 & -2 & 2 & -2 & 2 & -2 & -2 & 2 & -2 & 2 & -2 & 2 & -2 & -2 \\
    1 & 1 & 1 & 1 & 1 & 1 & 1 & 1 & 2 & 2 & -2 & 2 & -2 & -2 & -2 & 2 & 2 & -2 & 2 & -2 & -2 & -2 \\
    1 & 1 & 1 & 1 & 1 & 1 & 1 & 1 & -2 & -2 & 2 & 2 & -2 & -2 & 2 & -2 & -2 & 2 & 2 & -2 & -2 & 2 \\
    1 & 1 & 1 & 1 & 1 & 1 & 1 & 1 & -2 & 2 & -2 & -2 & 2 & -2 & 2 & -2 & 2 & -2 & -2 & 2 & -2 & 2 \\
    1 & 1 & 1 & 1 & 1 & 1 & 1 & 1 & 2 & -2 & -2 & -2 & -2 & 2 & 2 & 2 & -2 & -2 & -2 & -2 & 2 & 2 \\
    1 & 1 & 1 & 1 & 1 & 1 & 1 & 1 & -2 & -2 & -2 & 2 & 2 & 2 & -2 & -2 & -2 & -2 & 2 & 2 & 2 & -2 \\
    2 & -2 & 2 & 2 & -2 & -2 & 2 & -2 & 4 & 0 & 0 & 0 & 0 & 0 & 0 & -4 & 0 & 0 & 0 & 0 & 0 & 0 \\
    2 & 2 & -2 & 2 & -2 & 2 & -2 & -2 & 0 & 4 & 0 & 0 & 0 & 0 & 0 & 0 & -4 & 0 & 0 & 0 & 0 & 0 \\
    2 & 2 & 2 & -2 & 2 & -2 & -2 & -2 & 0 & 0 & 4 & 0 & 0 & 0 & 0 & 0 & 0 & -4 & 0 & 0 & 0 & 0 \\
    2 & -2 & -2 & 2 & 2 & -2 & -2 & 2 & 0 & 0 & 0 & 4 & 0 & 0 & 0 & 0 & 0 & 0 & -4 & 0 & 0 & 0 \\
    2 & -2 & 2 & -2 & -2 & 2 & -2 & 2 & 0 & 0 & 0 & 0 & 4 & 0 & 0 & 0 & 0 & 0 & 0 & -4 & 0 & 0 \\
    2 & 2 & -2 & -2 & -2 & -2 & 2 & 2 & 0 & 0 & 0 & 0 & 0 & 4 & 0 & 0 & 0 & 0 & 0 & 0 & -4 & 0 \\
    2 & -2 & -2 & -2 & 2 & 2 & 2 & -2 & 0 & 0 & 0 & 0 & 0 & 0 & -4 & 0 & 0 & 0 & 0 & 0 & 0 & 4 \\
    2 & -2 & 2 & 2 & -2 & -2 & 2 & -2 & -4 & 0 & 0 & 0 & 0 & 0 & 0 & 4 & 0 & 0 & 0 & 0 & 0 & 0 \\
    2 & 2 & -2 & 2 & -2 & 2 & -2 & -2 & 0 & -4 & 0 & 0 & 0 & 0 & 0 & 0 & 4 & 0 & 0 & 0 & 0 & 0 \\
    2 & 2 & 2 & -2 & 2 & -2 & -2 & -2 & 0 & 0 & -4 & 0 & 0 & 0 & 0 & 0 & 0 & 4 & 0 & 0 & 0 & 0 \\
    2 & -2 & -2 & 2 & 2 & -2 & -2 & 2 & 0 & 0 & 0 & -4 & 0 & 0 & 0 & 0 & 0 & 0 & 4 & 0 & 0 & 0 \\
    2 & -2 & 2 & -2 & -2 & 2 & -2 & 2 & 0 & 0 & 0 & 0 & -4 & 0 & 0 & 0 & 0 & 0 & 0 & 4 & 0 & 0 \\
    2 & 2 & -2 & -2 & -2 & -2 & 2 & 2 & 0 & 0 & 0 & 0 & 0 & -4 & 0 & 0 & 0 & 0 & 0 & 0 & 4 & 0 \\
    2 & -2 & -2 & -2 & 2 & 2 & 2 & -2 & 0 & 0 & 0 & 0 & 0 & 0 & 4 & 0 & 0 & 0 & 0 & 0 & 0 & -4
  \end{array}\right).
\end{align*}
    
\twocolumngrid

    
11 types of gapped boundaries are:
\begin{align*}
  \begin{array}{cl}
    (1) \quad&  \cl{1}\opl\cl{2}\opl\cl{10}\opl\cl{11}\opl\cl{14}\, ,\\
    (2) \quad&  \cl{1}\opl\cl{3}\opl\cl{9}\opl\cl{11}\opl\cl{13} \,,\\
    (3) \quad&  \cl{1}\opl\cl{4}\opl\cl{9}\opl\cl{10}\opl\cl{12} \,,\\
    (4) \quad&  \cl{1}\opl\cl{8}\opl\cl{12}\opl\cl{13}\opl\cl{14}\, ,\\
    (5) \quad&  \cl{1}\opl\cl{2}\opl\cl{3}\opl\cl{5}\opl2 \cl{11}\, ,\\
    (6) \quad&  \cl{1}\opl\cl{2}\opl\cl{4}\opl\cl{6}\opl2 \cl{10}\, ,\\
    (7) \quad&  \cl{1}\opl\cl{2}\opl\cl{7}\opl\cl{8}\opl2 \cl{14}\, ,\\
    (8) \quad&  \cl{1}\opl\cl{3}\opl\cl{4}\opl\cl{7}\opl2 \cl{9} \,,\\
    (9) \quad&  \cl{1}\opl\cl{3}\opl\cl{6}\opl\cl{8}\opl2 \cl{13}\, ,\\
    (10)\quad&  \cl{1}\opl\cl{4}\opl\cl{5}\opl\cl{8}\opl2 \cl{12}\, ,\\
    (11)\quad&
      \cl{1}\opl\cl{2}\opl\cl{3}\opl\cl{4}\opl\cl{5}\opl\cl{6}\opl\cl{7}\opl\cl{8}\,.
  \end{array}
\end{align*}
    
GSD on a cylinder with two gapped boundaries are computed: 
\begin{align*}
  \begin{array}{c|ccccccccccc}
    \GSD    & (1)&(2)&(3)&(4)&(5)&(6)&(7)&(8)&(9)&(10)&(11)\\\hline
    (1) &  5&2&2&2&4&4&4&1&1&1&2\\
    (2) &  2&5&2&2&4&1&1&4&4&1&2\\
    (3) &  2&2&5&2&1&4&1&4&1&4&2\\
    (4) &  2&2&2&5&1&1&4&1&4&4&2\\
    (5) &  4&4&1&1&8&2&2&2&2&2&4\\
    (6) &  4&1&4&1&2&8&2&2&2&2&4\\
    (7) &  4&1&1&4&2&2&8&2&2&2&4\\
    (8) &  1&4&4&1&2&2&2&8&2&2&4\\
    (9) &  1&4&1&4&2&2&2&2&8&2&4\\
    (10)&  1&1&4&4&2&2&2&2&2&8&4\\
    (11)&  2&2&2&2&4&4&4&4&4&4&8
  \end{array}
\end{align*} 

\onecolumngrid


\newpage
\subsection{XII. Gapped boundaries of $D(Q_8)$ phase: 6 types} 

Note that $D^{}(Q_8) = D^{\omega_3[{3i}]}(Z_2{}^3) =D^{\alpha_1}(D_4) = D^{\alpha_2}(D_4)$.
The $\cS,\cT$ matrices of $D(Q_8)$ are~\cite{Propitius95,WangWen1404}:


  \begin{align*}
    \cT&=\Diag(1,1,1,1,1,1,1,1,\ii,\ii,\ii,-1,-1,-1,1,-\ii,-\ii,-\ii,1,1,1,-1),
    \\
    \cS&=\frac{1}{8}\left(
    \begin{array}{cccccccccccccccccccccc}
      1 & 1 & 1 & 1 & 1 & 1 & 1 & 1 & 2 & 2 & 2 & 2 & 2 & 2 & 2 & 2 & 2 & 2 & 2 & 2 & 2 & 2 \\
      1 & 1 & 1 & 1 & 1 & 1 & 1 & 1 & -2 & 2 & 2 & -2 & -2 & 2 & -2 & -2 & 2 & 2 & -2 & -2 & 2 & -2 \\
      1 & 1 & 1 & 1 & 1 & 1 & 1 & 1 & 2 & -2 & 2 & -2 & 2 & -2 & -2 & 2 & -2 & 2 & -2 & 2 & -2 & -2 \\
      1 & 1 & 1 & 1 & 1 & 1 & 1 & 1 & 2 & 2 & -2 & 2 & -2 & -2 & -2 & 2 & 2 & -2 & 2 & -2 & -2 & -2 \\
      1 & 1 & 1 & 1 & 1 & 1 & 1 & 1 & -2 & -2 & 2 & 2 & -2 & -2 & 2 & -2 & -2 & 2 & 2 & -2 & -2 & 2 \\
      1 & 1 & 1 & 1 & 1 & 1 & 1 & 1 & -2 & 2 & -2 & -2 & 2 & -2 & 2 & -2 & 2 & -2 & -2 & 2 & -2 & 2 \\
      1 & 1 & 1 & 1 & 1 & 1 & 1 & 1 & 2 & -2 & -2 & -2 & -2 & 2 & 2 & 2 & -2 & -2 & -2 & -2 & 2 & 2 \\
      1 & 1 & 1 & 1 & 1 & 1 & 1 & 1 & -2 & -2 & -2 & 2 & 2 & 2 & -2 & -2 & -2 & -2 & 2 & 2 & 2 & -2 \\
      2 & -2 & 2 & 2 & -2 & -2 & 2 & -2 & -4 & 0 & 0 & 0 & 0 & 0 & 0 & 4 & 0 & 0 & 0 & 0 & 0 & 0 \\
      2 & 2 & -2 & 2 & -2 & 2 & -2 & -2 & 0 & -4 & 0 & 0 & 0 & 0 & 0 & 0 & 4 & 0 & 0 & 0 & 0 & 0 \\
      2 & 2 & 2 & -2 & 2 & -2 & -2 & -2 & 0 & 0 & -4 & 0 & 0 & 0 & 0 & 0 & 0 & 4 & 0 & 0 & 0 & 0 \\
      2 & -2 & -2 & 2 & 2 & -2 & -2 & 2 & 0 & 0 & 0 & 4 & 0 & 0 & 0 & 0 & 0 & 0 & -4 & 0 & 0 & 0 \\
      2 & -2 & 2 & -2 & -2 & 2 & -2 & 2 & 0 & 0 & 0 & 0 & 4 & 0 & 0 & 0 & 0 & 0 & 0 & -4 & 0 & 0 \\
      2 & 2 & -2 & -2 & -2 & -2 & 2 & 2 & 0 & 0 & 0 & 0 & 0 & 4 & 0 & 0 & 0 & 0 & 0 & 0 & -4 & 0 \\
      2 & -2 & -2 & -2 & 2 & 2 & 2 & -2 & 0 & 0 & 0 & 0 & 0 & 0 & 4 & 0 & 0 & 0 & 0 & 0 & 0 & -4 \\
      2 & -2 & 2 & 2 & -2 & -2 & 2 & -2 & 4 & 0 & 0 & 0 & 0 & 0 & 0 & -4 & 0 & 0 & 0 & 0 & 0 & 0 \\
      2 & 2 & -2 & 2 & -2 & 2 & -2 & -2 & 0 & 4 & 0 & 0 & 0 & 0 & 0 & 0 & -4 & 0 & 0 & 0 & 0 & 0 \\
      2 & 2 & 2 & -2 & 2 & -2 & -2 & -2 & 0 & 0 & 4 & 0 & 0 & 0 & 0 & 0 & 0 & -4 & 0 & 0 & 0 & 0 \\
      2 & -2 & -2 & 2 & 2 & -2 & -2 & 2 & 0 & 0 & 0 & -4 & 0 & 0 & 0 & 0 & 0 & 0 & 4 & 0 & 0 & 0 \\
      2 & -2 & 2 & -2 & -2 & 2 & -2 & 2 & 0 & 0 & 0 & 0 & -4 & 0 & 0 & 0 & 0 & 0 & 0 & 4 & 0 & 0 \\
      2 & 2 & -2 & -2 & -2 & -2 & 2 & 2 & 0 & 0 & 0 & 0 & 0 & -4 & 0 & 0 & 0 & 0 & 0 & 0 & 4 & 0 \\
      2 & -2 & -2 & -2 & 2 & 2 & 2 & -2 & 0 & 0 & 0 & 0 & 0 & 0 & -4 & 0 & 0 & 0 & 0 & 0 & 0 & 4 
    \end{array}\right).
  \end{align*}
  

There are 6 types of gapped boundaries:
\begin{align*}
 (1) \quad&\cl{1}\opl\cl{8}\opl\cl{19}\opl\cl{20}\opl\cl{21} \,,\\
 (2) \quad&\cl{1}\opl\cl{2}\opl\cl{7}\opl\cl{8}\opl2 \cl{21} \,,\\
 (3) \quad&\cl{1}\opl\cl{3}\opl\cl{6}\opl\cl{8}\opl2 \cl{20} \,,\\
 (4) \quad&\cl{1}\opl\cl{4}\opl\cl{5}\opl\cl{8}\opl2 \cl{19} \,,\\
 (5) \quad&\cl{1}\opl\cl{5}\opl\cl{6}\opl\cl{7}\opl2 \cl{15} \,,\\
 (6) \quad &\cl{1}\opl\cl{2}\opl\cl{3}\opl\cl{4}\opl\cl{5}\opl\cl{6}\opl\cl{7}\opl\cl{8}\,.
\end{align*}
GSD on a cylinder with two gapped boundaries are computed:
\begin{align*}
  \begin{array}{c|cccccc}
     \GSD   & (1)&(2)&(3)&(4)&(5)&(6)\\\hline
   (1) & 5&4&4&4&1&2\\
   (2) & 4&8&2&2&2&4\\
   (3) & 4&2&8&2&2&4\\
   (4) & 4&2&2&8&2&4\\
   (5) & 1&2&2&2&8&4\\
   (6) & 2&4&4&4&4&8
  \end{array}
\end{align*}


\newpage
\subsection{XIII. Gapped boundaries of $D^{\omega_3{[3d]}}(Z_2{}^3)$ phase: 5 types}

Note that $D^{\omega_3{[3d]}}(Z_2{}^3) = D^{\gamma^4}(Q_8)$.
The $\cS,\cT$ matrices of $D^{\omega_3{[3d]}}(Z_2{}^3)$ are~\cite{WangWen1404}:

  \begin{align*}
    \cT&=\Diag(1,1,1,1,1,1,1,1,1,1,1,\ii,\ii,\ii,1,-1,-1,-1,-\ii,-\ii,-\ii,-1),
    \\
    \cS&= \frac{1}{8} \left(
    \begin{array}{cccccccccccccccccccccc}
      1 & 1 & 1 & 1 & 1 & 1 & 1 & 1 & 2 & 2 & 2 & 2 & 2 & 2 & 2 & 2 & 2 & 2 & 2 & 2 & 2 & 2 \\
      1 & 1 & 1 & 1 & 1 & 1 & 1 & 1 & -2 & 2 & 2 & -2 & -2 & 2 & -2 & -2 & 2 & 2 & -2 & -2 & 2 & -2 \\
      1 & 1 & 1 & 1 & 1 & 1 & 1 & 1 & 2 & -2 & 2 & -2 & 2 & -2 & -2 & 2 & -2 & 2 & -2 & 2 & -2 & -2 \\
      1 & 1 & 1 & 1 & 1 & 1 & 1 & 1 & 2 & 2 & -2 & 2 & -2 & -2 & -2 & 2 & 2 & -2 & 2 & -2 & -2 & -2 \\
      1 & 1 & 1 & 1 & 1 & 1 & 1 & 1 & -2 & -2 & 2 & 2 & -2 & -2 & 2 & -2 & -2 & 2 & 2 & -2 & -2 & 2 \\
      1 & 1 & 1 & 1 & 1 & 1 & 1 & 1 & -2 & 2 & -2 & -2 & 2 & -2 & 2 & -2 & 2 & -2 & -2 & 2 & -2 & 2 \\
      1 & 1 & 1 & 1 & 1 & 1 & 1 & 1 & 2 & -2 & -2 & -2 & -2 & 2 & 2 & 2 & -2 & -2 & -2 & -2 & 2 & 2 \\
      1 & 1 & 1 & 1 & 1 & 1 & 1 & 1 & -2 & -2 & -2 & 2 & 2 & 2 & -2 & -2 & -2 & -2 & 2 & 2 & 2 & -2 \\
      2 & -2 & 2 & 2 & -2 & -2 & 2 & -2 & 4 & 0 & 0 & 0 & 0 & 0 & 0 & -4 & 0 & 0 & 0 & 0 & 0 & 0 \\
      2 & 2 & -2 & 2 & -2 & 2 & -2 & -2 & 0 & 4 & 0 & 0 & 0 & 0 & 0 & 0 & -4 & 0 & 0 & 0 & 0 & 0 \\
      2 & 2 & 2 & -2 & 2 & -2 & -2 & -2 & 0 & 0 & 4 & 0 & 0 & 0 & 0 & 0 & 0 & -4 & 0 & 0 & 0 & 0 \\
      2 & -2 & -2 & 2 & 2 & -2 & -2 & 2 & 0 & 0 & 0 & -4 & 0 & 0 & 0 & 0 & 0 & 0 & 4 & 0 & 0 & 0 \\
      2 & -2 & 2 & -2 & -2 & 2 & -2 & 2 & 0 & 0 & 0 & 0 & -4 & 0 & 0 & 0 & 0 & 0 & 0 & 4 & 0 & 0 \\
      2 & 2 & -2 & -2 & -2 & -2 & 2 & 2 & 0 & 0 & 0 & 0 & 0 & -4 & 0 & 0 & 0 & 0 & 0 & 0 & 4 & 0 \\
      2 & -2 & -2 & -2 & 2 & 2 & 2 & -2 & 0 & 0 & 0 & 0 & 0 & 0 & 4 & 0 & 0 & 0 & 0 & 0 & 0 & -4 \\
      2 & -2 & 2 & 2 & -2 & -2 & 2 & -2 & -4 & 0 & 0 & 0 & 0 & 0 & 0 & 4 & 0 & 0 & 0 & 0 & 0 & 0 \\
      2 & 2 & -2 & 2 & -2 & 2 & -2 & -2 & 0 & -4 & 0 & 0 & 0 & 0 & 0 & 0 & 4 & 0 & 0 & 0 & 0 & 0 \\
      2 & 2 & 2 & -2 & 2 & -2 & -2 & -2 & 0 & 0 & -4 & 0 & 0 & 0 & 0 & 0 & 0 & 4 & 0 & 0 & 0 & 0 \\
      2 & -2 & -2 & 2 & 2 & -2 & -2 & 2 & 0 & 0 & 0 & 4 & 0 & 0 & 0 & 0 & 0 & 0 & -4 & 0 & 0 & 0 \\
      2 & -2 & 2 & -2 & -2 & 2 & -2 & 2 & 0 & 0 & 0 & 0 & 4 & 0 & 0 & 0 & 0 & 0 & 0 & -4 & 0 & 0 \\
      2 & 2 & -2 & -2 & -2 & -2 & 2 & 2 & 0 & 0 & 0 & 0 & 0 & 4 & 0 & 0 & 0 & 0 & 0 & 0 & -4 & 0 \\
      2 & -2 & -2 & -2 & 2 & 2 & 2 & -2 & 0 & 0 & 0 & 0 & 0 & 0 & -4 & 0 & 0 & 0 & 0 & 0 & 0 & 4 \\
    \end{array}
    \right).
  \end{align*}

There are 5 types of gapped boundaries:
\begin{align*}
 (1) \quad&\cl{1}\opl\cl{2}\opl\cl{3}\opl\cl{5}\opl2 \cl{11}\,, \\
 (2) \quad&\cl{1}\opl\cl{2}\opl\cl{4}\opl\cl{6}\opl2 \cl{10}\,, \\
 (3) \quad&\cl{1}\opl\cl{3}\opl\cl{4}\opl\cl{7}\opl2 \cl{9} \,, \\
 (4) \quad&\cl{1}\opl\cl{5}\opl\cl{6}\opl\cl{7}\opl2 \cl{15}\,, \\
 (5) \quad &\cl{1}\opl\cl{2}\opl\cl{3}\opl\cl{4}\opl\cl{5}\opl\cl{6}\opl\cl{7}\opl\cl{8}\,.
\end{align*}
GSD on a cylinder with two gapped boundaries are computed:
\begin{align*}
 \begin{array}{c|ccccc}
      \GSD  & (1)&(2)&(3)&(4)&(5)\\\hline
    (1) &8&2&2&2&4\\
    (2) &2&8&2&2&4\\
    (3) &2&2&8&2&4\\
    (4) &2&2&2&8&4\\
    (5) &4&4&4&4&8
  \end{array}
\end{align*}


\newpage
\subsection{XIV. Gapped boundaries of $D^{\omega_3{[5]}}(Z_2{}^3)$ phase: 3 types}

Note that $D^{\omega_3{[5]}}(Z_2{}^3)= D^{\alpha_1\alpha_2}(D_4)$.
The $\cS,\cT$ matrices of $D^{\omega_3{[5]}}(Z_2{}^3)$ are~\cite{WangWen1404}:

  \begin{align*}
    \cT&=\Diag(1,1,1,1,1,1,1,1,\ii,\ii,1,-1,\ii,\ii,-\ii,-\ii,-\ii,-1,1,-\ii,-\ii,\ii),
    \\  
    \cS&= \frac{1}{8} \left(
    \begin{array}{cccccccccccccccccccccc}
      1 & 1 & 1 & 1 & 1 & 1 & 1 & 1 & 2 & 2 & 2 & 2 & 2 & 2 & 2 & 2 & 2 & 2 & 2 & 2 & 2 & 2 \\
      1 & 1 & 1 & 1 & 1 & 1 & 1 & 1 & -2 & 2 & 2 & -2 & -2 & 2 & -2 & -2 & 2 & 2 & -2 & -2 & 2 & -2 \\
      1 & 1 & 1 & 1 & 1 & 1 & 1 & 1 & 2 & -2 & 2 & -2 & 2 & -2 & -2 & 2 & -2 & 2 & -2 & 2 & -2 & -2 \\
      1 & 1 & 1 & 1 & 1 & 1 & 1 & 1 & 2 & 2 & -2 & 2 & -2 & -2 & -2 & 2 & 2 & -2 & 2 & -2 & -2 & -2 \\
      1 & 1 & 1 & 1 & 1 & 1 & 1 & 1 & -2 & -2 & 2 & 2 & -2 & -2 & 2 & -2 & -2 & 2 & 2 & -2 & -2 & 2 \\
      1 & 1 & 1 & 1 & 1 & 1 & 1 & 1 & -2 & 2 & -2 & -2 & 2 & -2 & 2 & -2 & 2 & -2 & -2 & 2 & -2 & 2 \\
      1 & 1 & 1 & 1 & 1 & 1 & 1 & 1 & 2 & -2 & -2 & -2 & -2 & 2 & 2 & 2 & -2 & -2 & -2 & -2 & 2 & 2 \\
      1 & 1 & 1 & 1 & 1 & 1 & 1 & 1 & -2 & -2 & -2 & 2 & 2 & 2 & -2 & -2 & -2 & -2 & 2 & 2 & 2 & -2 \\
      2 & -2 & 2 & 2 & -2 & -2 & 2 & -2 & -4 & 0 & 0 & 0 & 0 & 0 & 0 & 4 & 0 & 0 & 0 & 0 & 0 & 0 \\
      2 & 2 & -2 & 2 & -2 & 2 & -2 & -2 & 0 & -4 & 0 & 0 & 0 & 0 & 0 & 0 & 4 & 0 & 0 & 0 & 0 & 0 \\
      2 & 2 & 2 & -2 & 2 & -2 & -2 & -2 & 0 & 0 & 4 & 0 & 0 & 0 & 0 & 0 & 0 & -4 & 0 & 0 & 0 & 0 \\
      2 & -2 & -2 & 2 & 2 & -2 & -2 & 2 & 0 & 0 & 0 & 4 & 0 & 0 & 0 & 0 & 0 & 0 & -4 & 0 & 0 & 0 \\
      2 & -2 & 2 & -2 & -2 & 2 & -2 & 2 & 0 & 0 & 0 & 0 & -4 & 0 & 0 & 0 & 0 & 0 & 0 & 4 & 0 & 0 \\
      2 & 2 & -2 & -2 & -2 & -2 & 2 & 2 & 0 & 0 & 0 & 0 & 0 & -4 & 0 & 0 & 0 & 0 & 0 & 0 & 4 & 0 \\
      2 & -2 & -2 & -2 & 2 & 2 & 2 & -2 & 0 & 0 & 0 & 0 & 0 & 0 & -4 & 0 & 0 & 0 & 0 & 0 & 0 & 4 \\
      2 & -2 & 2 & 2 & -2 & -2 & 2 & -2 & 4 & 0 & 0 & 0 & 0 & 0 & 0 & -4 & 0 & 0 & 0 & 0 & 0 & 0 \\
      2 & 2 & -2 & 2 & -2 & 2 & -2 & -2 & 0 & 4 & 0 & 0 & 0 & 0 & 0 & 0 & -4 & 0 & 0 & 0 & 0 & 0 \\
      2 & 2 & 2 & -2 & 2 & -2 & -2 & -2 & 0 & 0 & -4 & 0 & 0 & 0 & 0 & 0 & 0 & 4 & 0 & 0 & 0 & 0 \\
      2 & -2 & -2 & 2 & 2 & -2 & -2 & 2 & 0 & 0 & 0 & -4 & 0 & 0 & 0 & 0 & 0 & 0 & 4 & 0 & 0 & 0 \\
      2 & -2 & 2 & -2 & -2 & 2 & -2 & 2 & 0 & 0 & 0 & 0 & 4 & 0 & 0 & 0 & 0 & 0 & 0 & -4 & 0 & 0 \\
      2 & 2 & -2 & -2 & -2 & -2 & 2 & 2 & 0 & 0 & 0 & 0 & 0 & 4 & 0 & 0 & 0 & 0 & 0 & 0 & -4 & 0 \\
      2 & -2 & -2 & -2 & 2 & 2 & 2 & -2 & 0 & 0 & 0 & 0 & 0 & 0 & 4 & 0 & 0 & 0 & 0 & 0 & 0 & -4 \\
    \end{array}
    \right).
  \end{align*}

There are 3 types of gapped boundaries:
\begin{align*}
(1) \quad& \cl{1}\opl\cl{2}\opl\cl{3}\opl\cl{5}\opl2 \cl{11} \,,\\
(2) \quad& \cl{1}\opl\cl{4}\opl\cl{5}\opl\cl{8}\opl2 \cl{19} \,,\\
(3) \quad& \cl{1}\opl\cl{2}\opl\cl{3}\opl\cl{4}\opl\cl{5}\opl\cl{6}\opl\cl{7}\opl\cl{8}
\,.
\end{align*}
GSD on a cylinder  with two gapped boundaries are computed:
\begin{align*}
  \begin{array}{c|ccc}
     \GSD   & (1)&(2)&(3)\\\hline
    (1) &8&2&4\\
    (2) &2&8&4\\
    (3) &4&4&8
  \end{array}
\end{align*}


\newpage
\subsection{XV. Gapped boundary of $D^{\omega_3{[7]}}(Z_2{}^3)$ phase: 1 type}

The $\cS,\cT$ matrices of $D^{\omega_3{[7]}}(Z_2{}^3)$ are~\cite{WangWen1404}:
  
  \begin{align*}
    \cT&=\Diag(1,1,1,1,1,1,1,1,\ii,\ii,\ii,-\ii,-\ii,-\ii,-\ii,-\ii,-\ii,-\ii,\ii,\ii,\ii,\ii),
    \\
    \cS&=
    \frac{1}{8} \left(
    \begin{array}{cccccccccccccccccccccc}
      1 & 1 & 1 & 1 & 1 & 1 & 1 & 1 & 2 & 2 & 2 & 2 & 2 & 2 & 2 & 2 & 2 & 2 & 2 & 2 & 2 & 2 \\
      1 & 1 & 1 & 1 & 1 & 1 & 1 & 1 & -2 & 2 & 2 & -2 & -2 & 2 & -2 & -2 & 2 & 2 & -2 & -2 & 2 & -2 \\
      1 & 1 & 1 & 1 & 1 & 1 & 1 & 1 & 2 & -2 & 2 & -2 & 2 & -2 & -2 & 2 & -2 & 2 & -2 & 2 & -2 & -2 \\
      1 & 1 & 1 & 1 & 1 & 1 & 1 & 1 & 2 & 2 & -2 & 2 & -2 & -2 & -2 & 2 & 2 & -2 & 2 & -2 & -2 & -2 \\
      1 & 1 & 1 & 1 & 1 & 1 & 1 & 1 & -2 & -2 & 2 & 2 & -2 & -2 & 2 & -2 & -2 & 2 & 2 & -2 & -2 & 2 \\
      1 & 1 & 1 & 1 & 1 & 1 & 1 & 1 & -2 & 2 & -2 & -2 & 2 & -2 & 2 & -2 & 2 & -2 & -2 & 2 & -2 & 2 \\
      1 & 1 & 1 & 1 & 1 & 1 & 1 & 1 & 2 & -2 & -2 & -2 & -2 & 2 & 2 & 2 & -2 & -2 & -2 & -2 & 2 & 2 \\
      1 & 1 & 1 & 1 & 1 & 1 & 1 & 1 & -2 & -2 & -2 & 2 & 2 & 2 & -2 & -2 & -2 & -2 & 2 & 2 & 2 & -2 \\
      2 & -2 & 2 & 2 & -2 & -2 & 2 & -2 & -4 & 0 & 0 & 0 & 0 & 0 & 0 & 4 & 0 & 0 & 0 & 0 & 0 & 0 \\
      2 & 2 & -2 & 2 & -2 & 2 & -2 & -2 & 0 & -4 & 0 & 0 & 0 & 0 & 0 & 0 & 4 & 0 & 0 & 0 & 0 & 0 \\
      2 & 2 & 2 & -2 & 2 & -2 & -2 & -2 & 0 & 0 & -4 & 0 & 0 & 0 & 0 & 0 & 0 & 4 & 0 & 0 & 0 & 0 \\
      2 & -2 & -2 & 2 & 2 & -2 & -2 & 2 & 0 & 0 & 0 & -4 & 0 & 0 & 0 & 0 & 0 & 0 & 4 & 0 & 0 & 0 \\
      2 & -2 & 2 & -2 & -2 & 2 & -2 & 2 & 0 & 0 & 0 & 0 & -4 & 0 & 0 & 0 & 0 & 0 & 0 & 4 & 0 & 0 \\
      2 & 2 & -2 & -2 & -2 & -2 & 2 & 2 & 0 & 0 & 0 & 0 & 0 & -4 & 0 & 0 & 0 & 0 & 0 & 0 & 4 & 0 \\
      2 & -2 & -2 & -2 & 2 & 2 & 2 & -2 & 0 & 0 & 0 & 0 & 0 & 0 & -4 & 0 & 0 & 0 & 0 & 0 & 0 & 4 \\
      2 & -2 & 2 & 2 & -2 & -2 & 2 & -2 & 4 & 0 & 0 & 0 & 0 & 0 & 0 & -4 & 0 & 0 & 0 & 0 & 0 & 0 \\
      2 & 2 & -2 & 2 & -2 & 2 & -2 & -2 & 0 & 4 & 0 & 0 & 0 & 0 & 0 & 0 & -4 & 0 & 0 & 0 & 0 & 0 \\
      2 & 2 & 2 & -2 & 2 & -2 & -2 & -2 & 0 & 0 & 4 & 0 & 0 & 0 & 0 & 0 & 0 & -4 & 0 & 0 & 0 & 0 \\
      2 & -2 & -2 & 2 & 2 & -2 & -2 & 2 & 0 & 0 & 0 & 4 & 0 & 0 & 0 & 0 & 0 & 0 & -4 & 0 & 0 & 0 \\
      2 & -2 & 2 & -2 & -2 & 2 & -2 & 2 & 0 & 0 & 0 & 0 & 4 & 0 & 0 & 0 & 0 & 0 & 0 & -4 & 0 & 0 \\
      2 & 2 & -2 & -2 & -2 & -2 & 2 & 2 & 0 & 0 & 0 & 0 & 0 & 4 & 0 & 0 & 0 & 0 & 0 & 0 & -4 & 0 \\
      2 & -2 & -2 & -2 & 2 & 2 & 2 & -2 & 0 & 0 & 0 & 0 & 0 & 0 & 4 & 0 & 0 & 0 & 0 & 0 & 0 & -4 \\
    \end{array}
    \right).
  \end{align*}
  

Only one type of gapped boundary is allowed:
\begin{align*}
& \cl{1}\opl\cl{2}\opl\cl{3}\opl\cl{4}\opl\cl{5}\opl\cl{6}\opl\cl{7}\opl\cl{8}\,.
\end{align*}
GSD on a cylinder  with two gapped boundaries of the same type must be 8.

\end{document}